
\input harvmac
\input tables
\def\np#1#2#3{Nucl. Phys. B{#1} (#2) #3}
\def\pl#1#2#3{Phys. Lett. {#1}B (#2) #3}

\Title{hep-th/yymmnn RU-95-51 SCIPP 95/41}
{\vbox{\centerline{Quantum Moduli Spaces of $N=1$ String Theories}}}
\bigskip
\centerline{Tom Banks}
\smallskip
\centerline{\it Department of Physics and Astronomy}
\centerline{\it Rutgers University, Piscataway, NJ 08855-0849}
\smallskip
\centerline{Michael Dine}
\smallskip
\centerline{\it Santa Cruz Institute for Particle Physics}
\centerline{\it University of California, Santa Cruz, CA   95064}
\bigskip
\baselineskip 18pt
\noindent
Generically, string models with $N=1$ supersymmetry are not
expected to have moduli beyond perturbation theory; stringy
non-perturbative effects as well as low energy field-theoretic
phenomena such as gluino condensation will lift any flat directions.
In this note, we describe models where some subspace of the
moduli space survives non-perturbatively.  Discrete $R$ symmetries
forbid
any inherently stringy
effects, and dynamical considerations control the field-theoretic
effects.  The surviving subspace is a space of high symmetry; the
system is attracted to this subspace by a potential which we compute.
Models of this type may be useful for considerations of duality
and raise troubling cosmological questions about string theory.
Our considerations also suggest a mechanism
for fixing the expectation value of the dilaton.
\Date{8/95}

\newsec{Introduction}

Known string models possess flat directions, directions in
field space along which the potential vanishes exactly.
The subspace of field space on which the classical potential
vanishes is called the space of moduli.
One of the most important practical problems of string theory
is to determine where in the space of moduli the true vacuum
state of string theory lies.

  In supersymmetric
classical vacua, approximate symmetries of the
theory guarantee that the superpotential is not
renormalized to all orders of perturbation theory.  These symmetries
are not exact and so we
 would seem to know very little about possible non-perturbative
contributions to the potential;
we can not even be sure about their general size
in the limit of weak coupling.  It is often assumed that effects
visible in the low energy theory such as gluino condensation
play the dominant role in lifting these flat directions,
but in general this is only an assumption.  Indeed, it is believed
that string theoretic effects may be as large as
$e^{-a/g}$, far larger than field theory phenomena.

In \ref\coping{T. Banks and M. Dine, Phys. Rev. {\bf D50} (1994) 7454.},
it was shown
that in many instances, exact discrete symmetries can be used to
bound the size of possible stringy non-perturbative effects.
In this note,
we show that it is often possible to make even stronger statements.
In particular, discrete symmetries sometimes
permit one to argue that there are {\it no} stringy
non-perturbative effects which correct the
superpotential on some subspace of
moduli space.  In theories with (tree level) anomalous $U(1)$ factors
in the gauge group,
there can be even stronger constraints.
We will present examples of
vacua with discrete symmetries and anomalous $U(1)$'s in which one can
argue that both stringy and field theoretic corrections to the
superpotential
vanish.  This means that on a subspace of the
classical moduli space, there are exact, quantum moduli.
We will also describe the dynamics as one approaches this
subspace of the moduli space.
 In others cases we will show that although there is no
quantum moduli space, it is possible to
describe the dynamics, at weak coupling, completely in terms
of well-understood low energy phenomena.

By Stringy Non-Perturbative effects, we have in mind
effects which appear in the effective lagrangian
at a scale just below the string scale.  This lagrangian is highly
constrained by symmetries and holomorphy of the superpotential
and the gauge coupling function.  In models with anomalous $U(1)$'s,
the $U(1)$ is broken by comparatively light fields:  the dilaton and
some (set of) charged chiral matter fields.  Thus there is a range
of energies for which it is appropriate to keep these fields in the
lagrangian.  For these fields, the symmetries are extraordinarily
restrictive.  In many cases, they forbid any corrections to the
superpotential which would lift the classical vacuum degeneracy.
Thus any
superpotential must be a low energy phenomenon.

In particular, we will be able to realize the suggestion
of \ref\berkooz{
T. Banks, M. Berkooz, P.J. Steinhardt, RU-94-92 (1995)
hep-th-9501053.}
that there are isolated minima of the potential on moduli space
at which SUSY is unbroken and the superpotential vanishes.
In \berkooz\
it was argued that such points would be dynamically selected in
postinflationary cosmology, even if other supersymmetric points
with lower vacuum energy exist on moduli space.
It was suggested that this cosmological selection principle
might resolve the classical vacuum degeneracy and pick out
our world from the modular muck of string theory.  Unfortunately,
our examples show that this is not the case.  The vacua that
we find do not resemble the real world.  In the conclusions
we will explore the consequences of this observation, which we feel
may be quite profound.

\newsec{\bf Absence of Stringy Non-perturbative Corrections}

To see the power of discrete symmetries and holomorphy
in restricting the form of stringy non-perturbative corrections
to the superpotential, consider the Calabi-Yau manifold
defined by the vanishing of a quintic polynomial in $CP^4$.
There is a subspace of the corresponding moduli space where
the model exhibits a large discrete
symmetry\ref\gsw{M. Green, J. Schwarz and E.
Witten, {\it Superstring
Theory}, Cambridge University Press, Cambridge (1987).}.
This symmetric subspace is described
by the quintic
$P= \sum Z_i^5$,
which respects symmetries under which each coordinate, $Z_i$, is
multiplied by $\alpha = e^{2 \pi i \over 5}$.  In addition, there
is the symmetry of permutation of each of the $Z_i$'s. In general
these symmetries are  R symmetries.
Under odd permutations, the superpotential is odd.  Under
the transformation
\eqn\ztransform{Z_i \rightarrow \alpha^{n_i} Z_i}
the superpotential transforms as
\eqn\wtransform{W \rightarrow \prod \alpha^{4n_i} W.}

The classical moduli  of these theories are easy to describe.
There is a modulus, the Kahler modulus, which describes
the overall size of the internal space.  This state
is invariant under all of the discrete symmetries.  Then there are a set
of moduli associated with deformations of the complex structure.
These are in one to one correspondence with quintic polynomials, and
we will denote them by
${\cal M}(a_1,a_2,a_3,a_4,a_5)$
where the associated polynomial
is $\prod Z^{a_i}$.  These fields transform like the corresponding
polynomial.

We will consider, first, the
case of the $E_8 \times E_8$ heterotic string.  In this case,
the moduli are paired with $27$'s of $E_6$.
At the symmetric point in the moduli space,
the discrete $R$ symmetries
forbid terms in the superpotential of the form:
\eqn\wforbidden{W= f(S)~~~~~W={\cal M} f(S).}
Here ${\cal M}$ denotes any of the complex structure
moduli with vanishing expectation values at this point, and $S$ is
the dilaton.
So no stringy non-perturbative correction can lift the degeneracy!
Indeed, there are numerous other couplings which vanish classically
which cannot receive stringy corrections.  In particular, all of the
massless states of the theory remain massless at this level.
This statement applies both to the moduli and to the matter
fields in the $27$ and $\bar 27$ of $E_6$, as well as the
various $E_6$ singlets.

This is not to say that the vacuum degeneracy cannot be lifted.
Effects involving the light fields present in the low energy
theory can remove the degeneracy.  In the present case, the
principle effect is gluino condensation.  In  \coping, we explained
how the resulting superpotential is consistent with the symmetries.
The point is that light fields can spontaneously break such symmetries.
What is striking in this theory is that,
at least on this subspace of the
moduli space, we can predict the precise form of the superpotential.

These arguments do not significantly restrict the
form of corrections to the Kahler potential.  Most authors
neglect these corrections because they vanish at weak enough coupling.
Indeed, at strong coupling, the whole concept of a ``light sector''
in string theory becomes meaningless, so one might imagine that
the Kahler potential in the effective lagrangian for the light sector
fields would only be a sensible notion at weak coupling.  However,
we have argued in \coping\ that the corrections to the Kahler potential
might be large in a region where the field theoretic coupling is
weak and the
light spectrum is identical
to that at weak coupling.
Thus, for our purposes, an exact determination of the
superpotential
is not yet a complete description of the dynamics on moduli space.

Away from the symmetric point, with one mild assumption, one
can still significantly constrain the size of non-perturbative effects.
In \coping, it was pointed out that if stringy non-perturbative
effects can be described by two dimensional field theories (e.g.
string instantons), then there is an exact symmetry under which
the axion shifts by $2 \pi$.  This means that stringy non-perturbative
effects are necessarily proportional to $e^{-nS}$, where
$n$ is an integer.  So, for small coupling, {\it everywhere in the
moduli space}, gluino condensation remains by far the
most important effect in lifting the classical vacuum
degeneracy.   The arguments for this periodicity
are admittedly somewhat shaky (though they are probably on a firmer
footing than most arguments for $S$ duality, of which this
symmetry is a subgroup).  As a result, in ref. \coping, various
discrete gauge symmetries under which $S$ transforms non-linearly
were used to constrain the superpotential.  In the rest of this
paper, however, we will make the stronger assumption of $2 \pi$
periodicity.

\newsec{\bf An $N=1$ Model with Non-Perturbative Moduli}

If we consider the compactification of the $O(32)$ theory
on the same Calabi-Yau manifold, a number of new phenomena
arise.  First, classically, the theory has no ``hidden sector"
gauge group.  The low energy, $O(26) \times U(1)$ theory is
extremely non-asymptotically free.  Second, the dilaton
transforms under the $U(1)$ symmetry; this leads to an
anomaly cancellation of the Green-Schwarz type.   As a consequence
of this transformation law, there is a Fayet-Iliopoulos D term.
This term can be canceled, within perturbation theory,
by giving an expectation value to some combination of matter
fields.   We will see in this section that, beyond
perturbation theory this is not necessarily the case.
Apart from the assumption of an
exact symmetry of $2 \pi$ shifts of the axion,
we will also assume that all terms allowed by symmetries are
generated with coefficients ${\cal O}(1)$.\foot{Needless
to say, this is a somewhat tricky
assumption.  Already in field theory, we know cases where it is
not quite true.}  With these assumptions, one can argue:
\item{1.}
It is still possible to cancel the $D$ terms.  However the
moduli space is significantly smaller at the non-perturbative level.
\item{2.}
Throughout the moduli space, some of the matter fields
($26$'s) gain mass.  In fact, at a generic point,
all but one of these fields gain mass.  In these regions, one
can determine the superpotential for the moduli and dilaton.
However, at least on some subspaces, the
theory is non-asymptotically free.  On these, one can argue
that no effect whatsoever lifts the flat directions.  Thus
the complete non-perturbative string theory still possesses
moduli.
\item{3.}
One can determine how the moduli superpotential and potential
behave as one approaches these regions of higher symmetry.
In fact, one finds that the moduli are attracted to these regions.

Consider the structure of this
model in more detail.  At tree level,
there are $101$ ``generations" and
one ``antigeneration."  A generation consists of a $26$ with
charge $+1$, $\phi^+$ and a singlet of charge $-2$, $N^{-2}$.
The antigeneration contains fields $\phi^-$ and $N^{+2}$.
The $U(1)$ is anomalous.  The anomaly is canceled by
assigning the dilaton field, which we write as $e^{-S}$,
charge $200$, i.e.
\eqn\stransform{e^{-S} \rightarrow e^{200 \alpha i} e^{-S}.}
A Fayet-Iliopoulos term is generated in this theory.
The sign of this term is such that it can be canceled by
one of the fields, $N^{-2}$.

States charged under $O(26)\times U(1)$ are in
one to one correspondence with these moduli.  We will
denote the $26$'s as $\phi^+(a_1, \dots a_5)$ (these
are the partners of the complex structure moduli) and
$\phi^-$.  The former, again, transform like the corresponding
polynomials while the latter is invariant.
The singlets will be denoted as $N^{-2}(a_1, \dots a_5)$
and $N^{+2}$.  They transform like the corresponding moduli
times $(1,1,1,1,1)(-1)^P$.


As we noted above, in this theory, a
Fayet-Iliopoulos term is generated at one
loop order with sign
such that one of the singlets, $N^{-2}$, obtains
a VEV.  In perturbation theory, no superpotential
involving the $N^{-2}$'s alone is permitted.
Similarly, there can be no term linear in
$N^{+2}$, $\phi^{-1}$ or the $O(26) \times U(1)$
singlet fields, $E$ (of which more below) alone,
times powers of $N^{-2}$.  As a result, the condition
to find a cancellation of the $D$ term, and hence a supersymmetric
minimum, is simply
\eqn\dcondition{g^2\vert N_i^{-2} \vert^2 = \mu^2.}
Note that this statement holds throughout the moduli space.

The description here of the $D$ term cancellation is inherently
perturbative.  In considering
what may happen non-perturbatively,
it is important to note that the charge of the field which gets
a VEV is opposite to that of $e^{-S}$.  As a consequence,
terms of the form $e^{-S} N^r$ can appear non-perturbatively,
for some $r$.  Such terms can play a role in lifting flat directions
and/or breaking supersymmetry, and will be important in
our subsequent discussion.

This fact (that the field which
gets a vev has charge opposite in sign
to $e^{-S}$) is quite generic to vacua with anomalous $U(1)$
factors, as long as we stick to the perturbative kinetic term
for the dilaton.  The anomaly is canceled by writing the
dilaton Kahler potential as $ - ln(S + S^* + q V)$, where $V$ is the
vector
potential superfield and $q$ is the anomaly coefficient.  The sign
of the charge of $e^{-S}$ is determined by $q$.  The contribution
of charged chiral superfields to the D-term is determined
by their kinetic term and charge.  The overall sign is determined
by positivity of the chiral kinetic term.  Cancellation of
the D-term then fixes the sign of the charged field which gets
a VEV to be opposite to that of $e^{-S}$.  However, if as
suggested in \coping , the real world lies in a region where the
dilaton Kahler potential is not given by lowest order perturbation
theory, it is possible to reverse this sign.  The dilaton contribution
to the D term depends on the first derivative of it's Kahler
potential, while its kinetic term (whose sign is fixed by positivity)
depends on the second derivative.  Some implications of this
possibility will be discussed in section 5.

Non-perturbatively, there are three important changes in
the analysis of the superpotential.
First, since $e^{-nS}$ has $U(1)$ charge $200 \times n$,
there are now terms which can appear in the superpotential
at the high energy scale involving the $N^{-2}$ fields.
Second, the Kahler potential may be appreciably modified.
This, however, does not qualitatively alter the problem
of solving the $D$ and $F$ term conditions at weak
coupling.  Third, there
may in principal be important effects in the low energy
theory which can contribute to the superpotential.

At a generic point in the moduli space, there are no
discrete symmetries, and a superpotential is permitted
which would lift all of the flat directions.  For small
values of the moduli, this superpotential has the form
\eqn\ndegen{W_{N}= P_{100}(N,{\cal M}) e^{-S}+ {\cal O}\left ( e^{-2S}
\right )}
where $P_{100}$ is a polynomial of degree $100$
in the $N^{-2}$ fields, and with suitable powers of the
moduli to satisfy all of the discrete symmetry constraints.
However, for small coupling, we expect field theoretic effects,
to be much larger than \ndegen ; we will see that this is the case.

It is particularly interesting to consider the theory
near the symmetric point.  In order to have an
allowed term in the superpotential, we need to find combinations
of $N$'s and $e^{-nS}$ which transform under the
discrete symmetries as $(4,4,4,4,4)$.  Consider first
the field $N(1,1,1,1,1)$. This is a rather symmetric
choice.  $N^{100}e^{-S}$ is invariant under the gauge symmetry,
but does not transform properly under the $R$ symmetries
(it is invariant).
Similarly, there is no term of the form
$N^{\prime} N^{99} e^{-S}$ which transforms in the
correct way.  So with respect to
the high energy superpotential, $N(1,1,1,1,1) \ne 0$
is an exact flat direction, perturbatively and non-perturbatively.

However, at scales below $\langle N \rangle$, field theoretic effects can
generate a superpotential that lifts this flat direction.
Many of the $\phi$
fields gain mass at this scale -- so many that the low energy theory
is asymptotically free.  Indeed, $\phi$'s corresponding
to polynomials of the form
$(1,1,1,1,1)$, (one field), $(2,1,1,1,0)$ (and permutations, for a total
of $20$ fields), and $(2,2,1,0,0)$ (an additional 30 fields)
gain mass.  All together, then, $51$ fields gain mass, leaving
$50$ fields in the low energy theory.  An $O(26)$ gauge theory
with $50$ fields in the fundamental representation, is asymptotically
free.  One might, then, expect important susy
breaking effects could occur, similar to
gluino condensation.  On the other hand,
the low energy is a theory in which, including
only renormalizable terms, no superpotential is generated
non-perturbatively.
We will discuss later the question of whether
including non-renormalizable terms,
a superpotential is generated, and turn first
to a theory in which the low energy theory is not
asymptotically free.

There are also perturbative vacua
for which the low energy theory is not asymptotically
free.  These will lead to realizations of the scenario of \berkooz\
and to an exact moduli space of $N=1$ supersymmetric vacua of string
theory.
Consider, for example, $N(3,2,0,0,0)$.  As before, at the high energy
scale, there are no terms which are permitted in the
superpotential which can lift the flat direction.  To see this,
consider the conditions for a coupling to appear in the superpotential:
\item{1.}
For couplings of the form $N^r e^{-nS}$,
\eqn\nsupr{(4r,3r,r,r,r)=(4,4,4,4,4)}
\item{2.}
For couplings of the form, $N^r N^{\prime}(a,b,c,c,c) e^{-nS}$
\eqn\nnprime{ (4r+a+1,3r+b+1,r+c+1,r+c+1,r+c+1)=(4,4,4,4,4)}
and $r=-1$ (mod 5).  This has no solutions.
\item{3.}
For couplings of the form $E N^r e^{-nS}$ or $M N^4 e^{-nS}$
where $E$ are gauge singlet fields and $M$ are the perturbative
moduli, there are no solutions, since no single modulus or singlet
transforms correctly.  (One can obtain the transformation laws
of the singlets, for example, from the work of Gepner\ref\gepner{
D. Gepner, Phys. Lett. {\bf 199B} (1987) 380.}
\vskip.2in

The counting of fields which gain mass in this direction is not
complicated.
Looking at terms of the form
$N \phi \phi^{\prime}$ (terms with higher powers of $N$ and factors
of $e^{-nS}$ transform in the same way).  Without loss of generality,
we can take $\phi(0,1,a,b,c)$ and $\phi^{\prime}(0,0,a^{\prime},
b^{\prime} c^{\prime})$.  Now it is not difficult to enumerate
the possible solutions.  One can take $(a,b,c)=(2,2,0)$ (and perms),
$(a,b,c)=(2,1,1)$ (and perms).  To each choice, there corresponds
a unique $\phi^{\prime}$.  All together, then, $24$ fields gain
mass.  This leaves $77$ massless $26$'s in the low energy theory,
and the theory at low energies is not asymptotically free.

We have already remarked that
away from the symmetric point, there is in general no solution
of the $D$ and $F$ term conditions.
As a result, there is a potential for
the
moduli and the fields $N$.  This potential, however, is proportional
to $e^{-S}$, which is small compared to expected effects
in the low energy field theory.  So let us ignore this and
suppose that the $D$ term is canceled by an expectation value
for some set of $N^{-2}$'s.    Then, at a generic
point in the moduli space all of the $\phi^+$ fields can
gain mass except
$\phi^{-}$, since there are no longer discrete symmetries
which prevent the coupling of $N^{-2}$ to any combination
of $\phi$'s.  For small ${\cal M}$, one can think of these terms
as arising through terms in the superpotential of the form
\eqn\modulicouplings{\phi^+ \phi^+ N^{-2}
{\cal M}^n.}
Note that the arguments of ref. \ref\dg{J. Distler and B. Greene,
Nucl. Phys. {\bf B309} (1988) 295.} do not forbid such couplings.
Throughout our discussion, we will assume that they are present
whenever permitted by symmetries.
$N^+$ also gains mass, through terms of the form
\eqn\nplusmass{N^+ N^+ N^- N^- {\cal M}^4.}
Integrating out $N^+$, gives a coupling $(\phi^{-})^4$ (we will
work out the moduli dependence shortly).  As a result, the low
energy theory of the matter fields possesses no flat directions.
A superpotential is generated, and there are minima with
unbroken supersymmetry.  At the minimum, the superpotential
has a non-vanishing expectation value.  Since the parameters of
this low
energy theory depend upon the moduli, this corresponds to the
generation of a superpotential for the moduli.

The unique superpotential which respects the non-anomalous
$R$ symmetry (which exists when non-renormalizable couplings
are ignored)  behaves as
\eqn\wotwentysix{W_{np}= {(\Lambda)^{71/23} \over
(\phi^2)^{1/23}}.}
Here, $\Lambda$ is the scale of the $O(26)$ theory
in which the fermions are massive.  It is related to $M$, the
string scale, and the fermion masses, by
\eqn\lambdaeqn{\Lambda = M e^{-S/b_o^{\prime}}
\left ({\det(m_f /M)}
\right )^{1 \over b_o^{\prime}}.}
Again, the primed quantities refer to the theory with massive
fermions; the unprimed quantities to the theory with all fermions
massless.  In this case,
\eqn\lambdatwo{\Lambda= Me^{-S/71} \left ({\det(m_f / M)}
\right )^{1/71}.}

For weak coupling, this is far larger than $e^{-S}$.  It is natural
to ask what happened to the shift symmetry.  This was explained
in ref. \coping.  The point is that the low energy theory has an
approximate $Z_{142}$ discrete symmetry, which is spontaneously
broken by the $\phi$ VEV.  Shifts in $a$ correspond to changes
of this VEV by a discrete phase.  Such phenomena can occur
in the low energy theory, where even at weak coupling,
light fields can gain VEV's and break symmetries.  It is also
natural to pause and ask what happened to
the $U(1)$ gauge symmetry in this expression? After all, one
might have been somewhat uneasy about using a broken
symmetry to constrain the form of an effective lagrangian.
However, this expression does conserve
the $U(1)$ once one takes account of the $N$ dependence
of $\Lambda$.  This $N$ dependence can be
determined from the renormalization group, and holomorphy:
\eqn\lambdathree{\Lambda = Me^{-{8\pi^2 \over b_o^{\prime}
g(M)^2} +
{1 \over b_o^{\prime}}\sum \ln({m_i/M^2})}}
where $M$ denotes some large scale such as the Planck mass
or string scale, $b_o^{\prime}$ is the first term
in the low energy $\beta$ function, and $m_i$
are the masses of the fields which gain mass through
$<N>$.  Since $N_i \propto N$, one finds that $\Lambda \propto
N^{101/71}e^{-S/71}$, which means that it has $U(1)$
charge $-2/71$.  Thus the full superpotential,
$W_{np}$ is invariant.

Including the non-renormalizable term, then, the relevant
superpotential
is
\eqn\one{W={(\Lambda)^{71/23} \over (\phi^-)^{2/23}}
+ {1 \over M} (\phi^-)^4.}
A superpotential of the general form
\eqn\nonflat{W= {\Lambda^a \over \phi^b}+ {1 \over  M} \phi^4}
has a minimum at
\eqn\three{<W>= ( M^{-b} \Lambda^{4a})^{1\over 4+b}.}
In the present case, this is
\eqn\four{<W> = ( M^{-1} \Lambda^{142})^{1/47}.}
The masses, as we will discuss shortly,
are determined by the moduli. So is the parameter
$M$.  Correspondingly, $<W>$ is a function
of the moduli -- indeed, it {\it is} the superpotential
for the moduli.

To get some idea how this behaves, let's suppose that all
of the moduli
have comparable VEV's, and ask how the determinant depends
on the moduli.  Take, again, the case $N=(3,2,0,0,0)$, and
ask how many powers of moduli are required to give
mass to each field.  The light $\phi$ fields are of the
following types:
$$(3,2,0,0,0) [2]~~~~(3,1,1,0,0) ~(3)~~~~(3,0,2,0,0) ~(3)
{}~~~~(3,0,1,1,0) ~(3)$$
$$(2,3,0,0,0)~[2]~~~~(2,2,1,0,0)~(3)[2]
{}~~~~(2,1,2,0,0)~ (3) ~~~~(2,1,1,1,0) ~(3) $$
$$(1,2,2,0,0)~(3)
{}~~~~(2,0,2,1,0)~(6) ~~~~(2,0,1,1,1)~~~~(1,3,1,0,0)~(3)[2]$$
$$
(1,2,2,0,0) ~(3)[2] ~~~~(1,2,1,1,0) ~(3)[2] ~~~~(1,1,3,0,0) ~(3)
{}~~~~(1,1,2,1,0)~ (6) $$
$$(1,1,1,1,1) ~~~~(1,0,3,1,0) ~(6)
{}~~~~(1,0,2,1,1)~ (3)  ~~~~(1,0,2,20)~(3) $$
$$~~~~(0,3,2,0,0) ~(3) ~~~~(0,3,1,1,0) ~(3)
{}~~~~(0,2,3,0,0) ~(3)
$$
\eqn\phimasses{~~~~(0,2,2,1,0) ~(6) ~~~~(0,2,1,1,1) .}
(77 states in all).  Most of these fields can gain
mass at first order in the moduli.  If higher powers of moduli
are required, we have indicated the number
in square braces.
In each case, this number can be determined by the
order of the polynomial required to give the correct transformation
properties under the discrete symmetries.  E.g. for the first field,
one needs a polynomial of degree 15 (after combining $\phi \times N$).
This can be provided by one $\phi$ and two moduli.  An example
of such a coupling is
$\phi(3,2,0,0,0) \phi(2,2,1,0,0)N(3,2,0,0,0){\cal M}(0,2,2,1,0){\cal
M}(0,0,0,2,3).$
It follows that the determinant
of this mass matrix
is of order ${\cal M}^{91}$.  By the same reasoning, the mass for the
$N^{+2}$ field is third order in the moduli.  It arises from
terms of the form $N^{+2}N^{+2}N^{-2}N^{-2}{\cal M}^3$
while the coupling
$\phi^{-} \phi^{-} N^{+2}$ is fifth order in moduli.

Now to determine the superpotential for the moduli,
we can follow \ref\seiberg{
N. Seiberg, \pl{318}{1993}{469}, hep-th--9309335; for
a review, see also hep-th--9408013.}.
We can view $\Lambda$ in eqn. \lambdathree\ as a function of the
quark masses.  This function is analytic, so the expression is valid
both for large and small masses.  Near the symmetric
point, we know the dependence of the masses on the moduli, so
in this way we obtain, from eqn. \four, the superpotential
near the symmetric point.
The result of this analysis is that in eqn. \nonflat,
we have
\eqn\lambdafour{\Lambda \propto {\cal M}^{91/71}~~~
 M^{-1}= {\cal M}^7.}
so
\eqn\wvev{<W>= W({\cal M})={\cal M}^{189/47}.}
Note, in particular, that $\Lambda \rightarrow 0$
slower than ${\cal M}^2$, so indeed for small ${\cal M}$
the masses are less than $\Lambda$.

The structure of this superpotential (and of the resulting potential)
is quite interesting.  First, it is  striking that the system
is attracted to the symmetric subspace of the moduli space.
Second, the superpotential has a branch cut starting at the origin.
This result is consistent with the fact that at the symmetric point,
there is a very large number of massless states, and the theory
is not asymptotically free.  Consequently, the symmetries which we
used to show
that there were exact flat directions in the high energy effective
lagrangian
are not spontaneously broken in the low energy theory.  The flat
directions are exactly flat!

There is a further subtlety which we have ignored up to this point.
As ${\cal M} \rightarrow 0$, $\phi \sim {\cal M}^{-35/47}\rightarrow
\infty$.  This is consistent with the fact
that, to all orders of perturbation theory, there is a flat direction
in the ${\cal M} \rightarrow 0$ limit.  Indeed, in perturbation
theory, the discrete symmetries, combined
with the $U(1)$ gauge invariance, forbid any operator of the
form $\phi^{-~n} N^{-2~m}$ or $\phi^{-~n} \phi^{\prime} N^{-2~m}$,
etc.  However, non-perturbatively, there are operators which can
appear and lift the flat direction.  For example
\eqn\liftsphi{(\phi^-)^3 \phi(3,2,0,0,0)N(3,2,0,0,0)^{99}e^{-S}+
\quad perms}
is consistent with all of the symmetries.  It involves no factors of
the moduli.

Once the moduli are sufficiently small, this term will become
important.  However, its effects still tend to
zero rapidly as the moduli tend to zero.  For non-zero ${\cal M}$,
the field $\phi(3,2,0,0,0)$ has a mass of order ${\cal M}^2$,
but the coupling $\phi(3,2,0,0,0)^2$ is of order ${\cal M}^3$.
Integrating out $\phi(3,2,0,0,0)$, then, leads to a sixth order
term in the $\phi$ superpotential of the form:
\eqn\wsix{W_6= {1 \over {\cal M}}(\phi^-)^6 e^{-2S}.}

In the limit of very small ${\cal M}$, neglecting the $\phi^4$ term,
we can take the superpotential to be
\eqn\smallm{W={{\cal M}^{91/23} \over \phi^{2/23}} +
{\cal M}^{-1} \phi^6
e^{-2S}.}
Solving for the minimum,
\eqn\smallmmin{\phi \sim {\cal M}^{114/140}~~~~~
W \sim {\cal M}^{12626/3220}.}
So the potential still goes rapidly to zero, but not quite as rapidly
as before.

Finally, it is worth noting that absence of asymptotic freedom
is not an essential requirement for the vanishing of the
potential.  In the classical flat direction with $N(1,1,1,1,1)\ne 0$,
a repetition of the analyses above gives a non-perturbative
superpotential behaving as
$$W_{np} \propto {\cal M}^{108/47}$$
Thus, in this case as well,
the potential vanishes rather rapidly for small ${\cal M}$, and
the symmetric point is the minimum.

\subsec{\it Discussion}

We have thus exhibited what appears to be a quantum moduli space
of string theory.  Symmetries constrain the quantum mechanically
exact superpotential in the effective lagrangian below the string scale
to vanish on a subspace of moduli space.  Low energy field theory
effects, which might spontaneously break these symmetries are
absent
because the low energy theory is infrared free.

There is a possible loophole in this argument, which has been revealed
by recent work on $N=2$ supersymmetric string and
gauge theories\ref\strom{
A. Strominger, hepth-9504090 (1995); P.C. Argyres, M.R. Douglas,
\break IASSNS-HEP-95-31 (1995) hepth-9505062. }.  In such
theories one
occasionally find points in moduli space at which soliton states
become
massless, while their classical size remains smaller than their
Compton wavelength.
If the solitons are magnetically charged under a perturbatively
infrared free gauge theory, they will change the sign of the
$\beta$ function for weak coupling.  The infrared dynamics
will be driven to a strong coupling fixed point, which is
as yet poorly understood.  Can we rule out the occurrence of such
phenomena on our putative quantum moduli space?

In analyzing this question we will make the assumption that, for
four dimensional compactifications, such effects manifest themselves
as singularities in the coefficients of string perturbation theory.
Indeed, a massless magnetic soliton state should show up in the
vacuum polarization function of the gauge fields.
For the symmetric point in complex structure moduli space,
both at large radius, and at the Gepner point, there are no such
singularities.
Local holomorphy of gauge kinetic functions then shows us that there
are no such singularities in a finite radius ball (in the space of all
moduli including the string coupling) around the zero coupling
symmetric
point (at {\it e.g.} the Gepner radius).  Within this ball our previous
analysis is valid, and there are no contributions to the superpotential
on our quantum moduli space.  Since the superpotential is locally
holomorphic,
it vanishes everywhere on the quantum moduli space.  Note
that this does
not imply that there are no massless soliton points in moduli space.
We learn only that these points are not on the quantum moduli space
for sufficiently weak coupling, and that they do not effect the
superpotential
on the part of moduli space we have explored.  In other words,
our results about the vanishing superpotential can be analytically
continued through any massless soliton points.  The soliton mass
will be a holomorphic function of the moduli which might happen
to have
a zero on a subspace of moduli space with finite complex
codimension.
Probably, if such points exist, they define new branches of the
moduli space that connect onto the branch we have studied along
this
submanifold of zeroes.

 \newsec{\bf Directions in Which the System is Repelled From the
Symmetric Point}

So far, we have studied in some detail an example in which a subspace
of the classical moduli space survives in the full quantum theory.
This subspace is a space where the theory exhibits a high
degree of symmetry, and it is perhaps not surprising that, at least
for weak string coupling, we have been able to show
that it is
a domain of attraction in the moduli space.
In this section we consider
an example where the opposite is true.  At weak coupling, the
degeneracy is lifted by non-perturbative effects.  However, the
system is repelled from the region of high symmetry.

The model is the four generation version of the quintic in $CP^4$
discussed in ref. \gsw.   This model is obtained by modding
out the Calabi-Yau space we have discussed earlier by
two freely acting $Z_5$ symmetries.  The resulting model
has four ``generations" ($\phi^+$ and $N^{-2}$) and one
``antigeneration" ($\phi^-$ and $N^{+2}$).
Following ref. \gsw,
we denote the generations by $\psi_o,\psi_2,\psi_{-2},\psi_1$
and $\psi_{-1}$ (where $\psi$ can be either $\phi^+$ or
$N^{-2}$); we will denote the antigeneration by $\bar \psi$.
The model
possesses two discrete symmetries, referred to as $W$ and $T$.
Under $W$, $\psi_n \rightarrow \psi_{-n}$, while $\bar \psi$
is neutral.
Under $T$,  $\psi_n\rightarrow e^{2 \pi i n / 5} \psi_n$.
$W$ is an $R$ symmetry under which the superpotential is odd.
$T$ is an ordinary symmetry.

At weak coupling,
we can analyze this model in much the same way as we analyzed
the models above.  Again, in this theory, a $D$ term is generated
at one loop, with a sign such that it can be canceled by
an expectation value for $N^{+}$.  The field $e^{-S}$ now has
charge $4$ under the $U(1)$.   At the classical level, there is
no obstacle to giving expectation values to the $N$ fields
so as to cancel the $D$ term.  Non-perturbatively, the symmetries
permit additional terms in the high scale effective lagrangian
which would lift the degeneracy, such as
\eqn\liftdegen{(N_2 N_2 e^{-S})^5 + ( 2 \rightarrow -2)
+ (N_1 N_1 e^{-S})^5.}
However, no terms are permitted at this level which
lift the $N_o$ direction.

As we will shortly see, however, field theoretic effects are far
more important than the highly suppressed stringy effects considered
above.  Let us examine the structure of the classical superpotential.
As explained in \gsw, the non-vanishing cubic terms are
\eqn\cubicallowed{\psi_o^3, \psi_o\psi_2\psi_{-2},
\psi_2\psi_{-1}\psi_{-1},
\psi_{-2}\psi_1\psi_1,\psi_2\psi_2\psi_1,
\psi_{-2}\psi_{-2}\psi_{-1}.}
Now consider a particular classical direction, $<N_2> \ne 0.$
In this direction, all of the $\phi_n$'s gain mass at tree level.
$\phi^-$, however, remains massless.  So at low energies,
the theory has the structure of an $O(26)$ SUSY gauge
theory with a single $26$.  At a generic point in moduli
space, just as in our earlier study, there are no flat directions.
In this case, however, the coefficient $1 \over M$ in
the superpotential,
$W= {1 \over M}(\phi^{-})^4$
 is linear in the moduli.  Repeating the analysis
of the preceding section, a non-perturbative superpotential
is seen to be generated.  It has the same form
as eqn. \one.  However,
in this case, $\Lambda$ is independent of the moduli.
Again, the minimum is given by eqn. \four.
This corresponds to a superpotential for the moduli which behaves as
${\cal M}^{1/47}$.
The resulting potential blows up as one approaches the symmetric
point,
almost as fast as $1 \over \vert {\cal M} \vert^2$.

This behavior is particularly easy to understand in the present case.
At the symmetric point, no additional fields transforming under
$O(26)$ becomes massless; the non-perturbative superpotential
has the same structure as at the generic point.   Thus for the
effective theory at (or near) this point, the
effective superpotential for
$N^{+2}$, $\phi^-$ and ${\cal M}$ has the structure
\eqn\waf{W_{eff}=
{\Lambda^{71/23} \over (\phi^-)^{2/23}}+ (\phi^-)^2N^{+2}
N^{+2}{\cal M}.}

It is easy to see that this superpotential has no supersymmetric
minimum.
It is possible to lower the energy by letting ${\cal
M}$ become large; this is precisely the repulsion we found above.
However, for sufficiently large ${\cal M}$, our analysis breaks down.
Thus, it is likely that there are supersymmetric minima of the
potential on moduli space at a finite distance (in string units)
from the symmetric point.  Generically, the potential at such points
will be negative \berkooz , and the moduli will not come to rest
at these minima of the potential after a period of inflation.
Thus, it might be that this entire region of moduli space
is ruled out by cosmological considerations.  We cannot of course
rule out the possibility that far from the symmetric point there is
a nonsupersymmetric minimum with nonnegative cosmological
constant,
which would be an attractor for the dynamics of postinflationary
cosmology.

Other directions (VEV's for $N$) can be studied in a similar
way.  The analysis is, in some cases, more complicated because
there are more light fields, but the results are similar.

\newsec{Some Strange Possibilities}

So far, the spirit of our discussion has been to work at very weak
coupling
and to establish the existence of moduli by working perturbatively
in some of the (other) moduli.  In this way, we have now established
the existence of exact quantum moduli for this theory, corresponding
at weak coupling to the original dilaton and to the radial dilaton.

It is natural to ask what sorts of phenomena might occur as we move
in towards stronger coupling.  One interesting possibility is the
following.
The kinetic
term for the dilaton has the structure
\eqn\dilatonkahler{\int d^4 \theta K(S + S^{\dagger} + V).}
The $D$ term is just $\partial K \over \partial S.$  A priori, we know
of no reason why this might not vanish for some value of $S$,
call it $S_o$.\foot{Since the Kahler potential is
negative, convex and increasing
as $S\rightarrow\infty$, its derivative could only vanish at finite $S_0$
if its second derivative were to change sign.  This would imply a
negative kinetic energy for the dilaton were it not for the possibility
of kinetic mixing between the dilaton and other moduli.  We assume that
such mixing occurs in the following speculations.  We thank V.
Kaplunovsky for a discussion of this point.
}  At this point, the potential has the
structure:
\eqn\dpotential{V=(m_V s + \sum q_i \vert \phi_i \vert^2)^2.}
Here $m_V^2= {\partial^2 K \over \partial S^2}$ is the mass
of the vector meson, and
and $s= Re(S-S_o)$ (up to a rescaling to give
a canonical kinetic term).  There are several interesting features of
this result:
\item{1.} At $s=0$, the theory is supersymmetric.  The
dilaton is completely absorbed by the supersymmetric
Higgs mechanism.
\item{2.} For $s>0$, fields with one sign of the $U(1)$ charge
get expectation values; for $s<0$, fields with the other sign
do.

There are a number of possibilities for the physics at such
a point, depending on the particle content of the model.
Here we list some of them (we do not claim to know
string vacua which realize every one of these possibilities,
but we know of no general argument that such vacua
cannot exist).
\item{1.}  There are no flat directions classically besides
the dilaton.  In this case, at one loop, the dilaton is fixed
at $S_o$.  Supersymmetry may be unbroken in the low
energy theory or it might be broken; if it is broken,
the scale of the breaking is of order $e^{-S_o}$.  As
we have argued elsewhere, it is perfectly possible
that, even though perturbation theory is not good for
the Kahler potential, this number is small.
\item{2.}  There may be several fields with one sign of the
charge, such that, classically, the $D$ term is canceled.
Non-perturbatively,
terms of the form $e^{-S}N^r$ will lift these flat directions.
Thus, at weak (but non-zero) coupling, there will be no ground state.
However, at strong coupling, the dilaton and/or fields with
the opposite sign of the $U(1)$ charge (for which no non-perturbative
superpotential may be permitted since $e^{-S}$ has the wrong
sign of the charge)  may have expectation values such that
the $D$ term vanishes.  In this case we would have an analytic moduli
space for $Re\  S < S_o$ but no vacuum state for larger values of the
real part of $S$.  This would provide a counterexample to the generic
analyticity of supersymmetric moduli spaces.

Clearly one can go on to enumerate further possibilities,
and one can speculate on connections to recent developments
with duality symmetries.  We will leave this for future work,
and just note that the vanishing of the $D$ term could
well play an important role in determining the fate of the dilaton.

\newsec{\bf Conclusions}

In this paper we have found examples of supersymmetric
stringy vacuum
states with vanishing cosmological constant.  In \berkooz , such states
were suggested as the natural postinflationary ground states
of string theory
for the nondilatonic moduli.  There it was emphasized that the modular
potential in such states would leave the dilaton direction flat.
It was suggested that lower energy nonperturbative gauge
dynamics
could lift this degeneracy.  If this led to a stable vacuum with
vanishing cosmological constant, it would have to break SUSY.

The states that we have discovered do not have these properties.
They
have no SUSY breaking low energy dynamics.  Indeed we have
argued that
they have an exact quantum moduli space of supersymmetric
vacua with
vanishing cosmological constant.  Apart from the disappointment of
finding that the \lq\lq cosmological vacuum selection principle'' of
\berkooz\ does not uniquely select the real world from among the
myriad points in moduli space, the existence of these states poses
a problem of principle for string theory. Why is the world we see
around us not in one of these highly stable states\foot{It is not
even clear that there is an anthropic answer to this question.
Such an argument would depend on the postinflationary history of
a universe which asymptotes to one of these states and would quickly
become enmired in unanswerable questions about whether there can
be life forms
radically different from ourselves.  }?

Superstring theory has long been known to have a quantum moduli
space
of vacua which do not resemble the real world.  These are states
with extended spacetime SUSY.  It has seemed possible to imagine that
these states are topologically disconnected from the part of moduli
space
in which our world lies, and that some sort of nonperturbative
anomaly might
afflict one and not the other.  Alternatively, one might attempt
to rule these states out cosmologically, since the quantum effective
potential
vanishes identically on the moduli space of extended SUSY ground
states.
Thus, if they are truly disconnected from the $N = 1$ classical moduli
space,
they could never be the result of a period of inflationary expansion.

By contrast, the quantum moduli space that we have discovered
lies right in the middle of the parts of classical moduli space
where potentials are generated. They do not look terribly
different than the classical vacua which resemble the world we live
in.
  The only kind of cosmological argument that might favor
one over the other is the observation of
\ref\moore{T. Banks, M. Berkooz, S.H. Shenker,
G. Moore and P. Steinhardt, RU-95-93 (1995) hep-th-9503114},
that superstring inflation
requires mild fine tuning.  In order to have $\sim 100$ e-foldings
of inflation, the curvature at the maximum of the
potential has to be one or two orders of magnitude smaller than
one would have guessed on the basis of dimensional analysis.
Thus there may be few places on moduli space where inflation occurs,
and a world like our own (rather than some point in the
supersymmetric
quantum moduli space) might be selected by the accident that
it lies near one of these inflationary maxima.
{\it A priori} it seems no less likely for one of the exact
supersymmetric quantum
ground states to
lie near an inflationary maximum of the potential than does
our own world.
Thus, until we understand a lot more about the potential on
moduli space than
we do at present, this will  not be a terribly satisfactory
explanation of
the properties of our world.

The work of Susskind\ref\antirem{L. Susskind,
SU-ITP-95-1 (1995) hep-th-9502069.} suggests an alternate
way to understand
the instability of vacua with extended SUSY.
Susskind\ref\priv{L. Susskind,
{\it private communication}} has speculated that black hole
evaporation
can lead to a unitary S-matrix only in four dimensions.  If this is
the case, and if string theory is consistent, then in higher
dimensional ground states, the information loss paradox of black holes
can only be resolved by the existence of remnants.  But in \antirem\
Susskind argues that in a theory with remnants flat space is unstable
to decay into a gas of remnants
(perhaps with an enormously long lifetime).
Thus, higher dimensional vacua might be unstable.  Vacuum states
with extended SUSY are continuously connected (even quantum
mechanically)
onto flat higher dimensional vacua, and would thus be unstable as well.

While we are not claiming that this particular line of reasoning is
the resolution of the problem of stable supersymmetric vacua,
we are intrigued by the possibility that it raises.  The
resolution of the puzzles of black hole evaporation is a problem
in nonperturbative string theory, and may well involve intrinsically
stringy dynamics which is not captured by low
energy field theory\ref\suss{L.
Susskind, SU-ITP-94-33 (1994) hep-th-9409089.}.
The preceding paragraph suggests that the stability
of string ground states may also be a problem that cannot be
analyzed completely by low energy effective field theory.

If this is the case, then our proof of nonperturbative stability of
a family of $N=1$ string vacua is incomplete, and might well be
incorrect.
The problem of resolving the degeneracy of classical
string ground states
would not reduce to the determination of the quantum effective
potential on the classical moduli space, but would have to be addressed
with the full nonperturbative apparatus of string theory (presently
unknown).  Our discovery of exact (at the level of
analysis of effective field theory), four dimensional, $N=1$
supersymmetric
vacua of string theory is a disquieting indication that the
construction and nonperturbative solution of quantum string theory
may be a prerequisite to answering even the most basic
questions about
the phenomenology of the theory.

Another (perhaps academic) question that is raised by the existence of
a quantum moduli space of $N=1$ vacua is whether the methods of
\ref\fhsv{S. Ferrara, J.A. Harvey, A. Strominger,
and C. Vafa EFI-95-26 (1995) hep-th-9505162.} and
particularly\ref\kv{
S. Kachru and C. Vafa, HUTP-95-A016 (1995) hep-th 9505105.}
could be generalized to give an
exact computation of the gauge kinetic functions on the moduli space.
For general $N=1$ compactifications, there does not seem to be any
sensible extension of \fhsv\kv.  Superpotentials will be generated
in all directions in classical moduli space, and for
coupling (${g^2 \over
 4\pi}$) of order one, there will be no sensible separation between
the moduli and the massive fields of the theory.  For such
compactifications,
the notion of moduli is an approximate one, and exact analytic
computations would seem impossible\foot{We thank
N. Seiberg for discussions
about this point.}.  Another indication of this is that we generically
expect $N=1$ compactifications to contain strongly
coupled gauge sectors
which will generate terms in the effective action which are not periodic
in the model independent axion.  The method of \fhsv and \kv
was to map one
string theory on another in such a way that the coupling
constant of the
original theory became a geometrical modulus of the dual
theory.  Quantum
calculations were then reduced to tree level calculations.
But tree level
string calculations never \lq\lq spontaneously break'' the periodicity
of geometrical moduli in the way that strong coupling effects
such as gaugino condensation break the axion periodicity.\foot{The only
apparent loophole in this argument that duality cannot give
us an exact
computation of gaugino condensation is the possibility that the duality
mapping appropriate for $N=1$ maps the dilaton onto a branched cover
of the geometrical moduli space of tree level string theory.
Possibly
this loophole can be closed by considering how the map must vary as
a function of those moduli which break the low energy gauge
group to an
abelian subgroup.   }

On the other hand, no such arguments prevent us from generalizing
\fhsv\kv to the quantum moduli spaces of $N=1$
compactifications that we have
discovered.  It seems likely to us that these methods might lead
to exact calculations of gauge kinetic functions on subspaces of
these moduli spaces.

Finally, the models we have considered here provide a natural
arena in which to discuss questions of duality in $N=1$ theories.
Generically, we expect that the classical flat directions of
$N=1$ theories are lifted, and the discussion of duality
might become somewhat murky.  Models with exact
non-perturbative
moduli are a natural arena in which to explore such dualities.
Indeed, examples of $N=1$ dualities
which have been uncovered recently\ref\vafawitten{
C. Vafa and E. Witten, hep-th--9507050.}\ref\harveyetal{J.A.
Harvey, D.A. Lowe and A. Strominger, hep-th--9507168.}
seem to possess this feature.  On the other hand, it may not be
so easy to guess the duals of some of the models we have
considered here.  It is natural to speculate, for example, that
under $S$ duality, the low energy gauge theory limit of a particular
string vacuum should go over to its Seiberg dual
\ref\intriligator{N. Seiberg, \np{345}{1995}{129}, hep-th--9411149;
K. Intriligator and N. Seiberg, \np{444}{1995}{125},
hep-th--9503179.}.
In this paper, we considered at some length a model
with gauge group $O(26)$ and $N_f=77$.  According to \intriligator,
the dual
theory has gauge group $O(55)$.
It is not easy to see how such a gauge group can arise
in any string theory.

\centerline{{\bf Acknowledgements}}
\noindent
We thank K. Intriligator,
N. Seiberg, S. Shenker L. Susskind, S. Thomas
and E. Witten for conversations.  The work of M.D. was supported
in part by the U.S. Department of Energy.
The work of T. Banks was supported in
part by the Department of Energy under
grant$\# DE-FG0590ER40559$.

\listrefs
\bye
\end